\documentclass[aps,pre,twocolumn,floatfix,superscriptaddress,showpacs,showkeys]{revtex4-2}
\usepackage{epsf,amsmath,amssymb,verbatim,color,multirow,pifont,graphicx,cleveref,tabularx}
\usepackage{graphicx}
\usepackage{epstopdf}
\usepackage{ulem}

\begin{document}
\title{Exploring growing complex systems with higher-order interactions}
\author{Soo Min Oh}%
\affiliation{Wireless Information and Network Sciences Laboratory, Massachusetts Institute of Technology, Cambridge, MA 02139, USA}
\affiliation{Laboratory for Information and Decision Systems, Massachusetts Institute of Technology, Cambridge, MA 02139, USA}
\affiliation{Center for Theoretical Physics, Seoul National University, Seoul 08826, Korea}
\author{Yongsun Lee}%
\affiliation{Center for Theoretical Physics, Seoul National University, Seoul 08826, Korea}
\affiliation{Department of Physics and Astronomy, Seoul National University, Seoul, 08826, Korea}
\author{Byungnam Kahng}%
\email{bkahng@kentech.ac.kr}
\affiliation{Center for Complex Systems Studies and Institute for Grid Modernization, Korea Institute of Energy Technology, Naju 58330 Korea}%
\date{\today}
\begin{abstract}
A complex system with many interacting individuals can be represented by a network consisting of nodes and links representing individuals and pairwise interactions, respectively. However, real-world systems grow with time and include many higher-order interactions. Such systems with higher-order interactions can be well described by a simplicial complex (SC), which is a type of hypergraph, consisting of simplexes representing sets of multiple interacting nodes. Here, capturing the properties of growing real-world systems, we propose a growing random SC (GRSC) model where a node is added and a higher dimensional simplex is established among nodes at each time step. We then rigorously derive various percolation properties in the GRSC. Finally, we confirm that the system exhibits an infinite-order phase transition as higher-order interactions accelerate the growth of the system and result in the advanced emergence of a giant cluster. This work can pave the way for interpreting growing complex systems with higher-order interactions such as social, biological, brain, and technological systems.
\end{abstract}

\maketitle

\section{Introduction}

Complex systems, characterized by numerous interacting components, can be effectively represented by networks where nodes represent individuals and links represent interactions between them. Traditional network models, which focus on pairwise interactions, have been instrumental in understanding a wide range of phenomena from social dynamics to biological processes~\cite{barabasi_emergence_1999,newman_structure_2003,boccaletti_complex_2006}. However, these models fall short in capturing the full complexity of real-world systems, especially those that grow with time and involve higher-order interactions.

Higher-order interactions, where multiple components interact simultaneously, are prevalent in many natural and artificial systems. For instance, in social networks, interactions often occur in groups rather than pairs~\cite{granovetter_strength_1973,watts_collective_1998,newman_scaling_1999,moody_structure_2004,greening_higher-order_2015,bianconi_higher-order_2021}. In biological systems, protein interactions frequently involve complexes of multiple proteins~\cite{sole_model_2002,vazquez_modeling_2002,hartwell_molecular_1999,spirin_protein_2003,barabasi_network_2004,goh_graph_2005,kim_fractal_2010,ispolatov_duplication-divergence_2005,kim_infinite-order_2002,oh_suppression_2018,oh_discontinuous_2019}, and in coauthorship networks~\cite{redner_how_1998,newman_structure_2001,newman_coauthorship_2004,barabasi_evolution_2002,lee_complete_2010}, academic papers are typically coauthored by multiple researchers. Traditional network models, which only account for pairwise interactions, do not adequately represent such complex interactions. Therefore, a more sophisticated framework is required to model these systems accurately.

Simplicial complexes (SCs) offer a powerful mathematical tool for capturing higher-order interactions. An SC is a type of hypergraph composed of simplexes, which generalize the concept of edges to higher dimensions. A $d$-dimensional simplex ($d$-simplex) with an integer $d$ is a convex hull of $n+1$ points that can be depicted as a filled polytope. Moreover, the convex hull of any nonempty subset of a $d$-simplex is referred to as a face of the $d$-simplex. The face of the $d$-simplex has a dimension denoted by $m$, with an integer value $m \le d$, and each face is itself a simplex. Facets are a set of maximal dimensional faces of a given SC. An individual is represented by a node, which is a $0$-simplex, and an interaction among $d+1$ individuals is described by a $d$-simplex. For $d=1$, $d=2$, $d=3$, and $d=4$, $d$-simplexes are presented by a line, triangle, tetrahedron, and $5$-cell, respectively. This SC representation allows for a comprehensive representation of complex interactions and is particularly suitable for modeling systems where higher-order interactions play a crucial role. 
Several studies have demonstrated the effectiveness of SCs for interpreting coauthorship complexes~\cite{lee_homological_2021,lee_betweenness_2021,oh_emergence_2021}, brain structures~\cite{petri_homological_2014,giusti_twos_2016,sizemore_cliques_2018}, and social contagion processes~\cite{bianconi_interdisciplinary_2015,iacopini_simplicial_2019,jhun_simplicial_2019,de_arruda_social_2020}.
%
Moreover, we note that numerous real-world systems grow by the continuous addition of new nodes and interactions~\cite{dorogovtsev_structure_2000,callaway_are_2001,dorogovtsev_anomalous_2001,dorogovtsev_evolution_2002}, and this growth is also a fundamental aspect of their evolution and impacts their structural and dynamic properties.

In our previous study~\cite{lee_homological_2021,oh_emergence_2021}, we proposed a growing scale-free simplicial complex (GSFSC) model that captures both the growth and higher-order interactions in real-world systems. We considered the dimension of simplexes to be uniformly set to $2$. However, real-world systems include interactions of diverse dimensions. Here, in this work, we simplify the model into a growing random simplicial complex (GRSC) model for simplicity and generalize the model from $2$-simplexes to $d$-simplexes for any positive integer $d$.
The GRSC model introduced in this work reflects the natural growth observed in real-world systems and incorporates the complexity of higher-order interactions. 
The model starts with a set of isolated nodes. At each time step, a new node ($0$-simplex) is added to the system, and a $d$-simplex is established among $d+1$ nodes. This iterative process is designed to mimic the way real-world systems grow and evolve over time. By capturing both the growth and higher-order interactions, the GRSC model provides a more realistic and comprehensive framework for studying complex systems.

We first consider the simplest case of $2$-simplexes and then generalize the GRSC model to $d$-simplexes for any positive integer $d$. This generalization allows us to explore a wide range of systems with different types and dimensions of interactions. We rigorously derive the rate equations for cluster size distributions within the GRSC model and obtain analytical solutions for various properties, including giant cluster sizes and second-order moments. Our analysis reveals that the giant cluster emerges smoothly and the second-order moment exhibits a discontinuity with two distinct finite limit values around the percolation thresholds. In addition to cluster size distributions, we analytically derive the degree distributions for both graph and facet degrees, confirming that these distributions follow an exponential form. Our results show that the percolation thresholds occur earlier with increasing $d$, indicating that higher-order interactions among nodes accelerate the growth of the system. This accelerated growth leads to an infinite-order phase transition, a feature of the complex dynamics governed by higher-order interactions.

Our GRSC model not only enhances our understanding of the structural and dynamic properties of growing complex systems with higher-order interactions but also provides a robust framework for further exploration. We anticipate that our findings will pave the way for interpreting various real-world systems, such as social, biological, brain, and technological networks, through the lens of higher-order interactions. By laying a solid foundation for understanding these systems, we aim to contribute to the broader field of complex system research and offer a valuable framework for future studies. In summary, this work proposes and investigates a novel model that bridges the gap between the static representation of higher-order interactions and the dynamic nature of growing real-world systems. By offering a comprehensive analytical framework, we lay the groundwork for future studies aimed at unraveling the complexities of higher-order interactions in growing systems.

This paper is organized as follows: In Sec.~\ref{sec:Growing random simplicial complex (GRSC) model with 2-simplexes}, we introduce a GRSC model with a $2$-simplex established with time and develop rate equations of the cluster size distributions. Based on the generating function approaches, we find the analytical form of giant cluster sizes and second-order moments. Subsequently, we analytically derive the graph and facet degree distributions. In Sec.~\ref{sec:GRSC model with $d$-simplexes}, we generalize the proposed approaches to a GRSC model with a $d$-simplex established with time. Finally, we conclude our work in Sec.~\ref{sec:Conclusions}.

\section{Growing random simplicial complex (GRSC) model with $2$-simplexes}
\label{sec:Growing random simplicial complex (GRSC) model with 2-simplexes}

\subsection{Models and rate equations}
\label{subsec:Models and rate equations with 2-simplexes}

\begin{figure*}[!t]
\center
\includegraphics[width=1.0\linewidth]{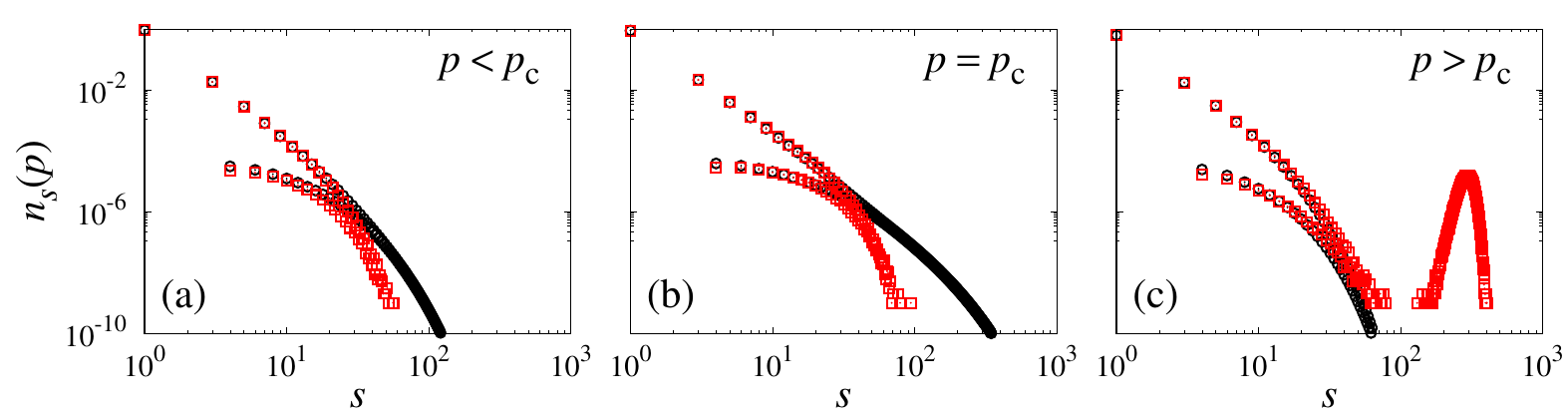}
\caption{(Color online)
Plots of the cluster size distribution $n_s(p)$ for (a) $p = 0.03 < p_c$, (b) $p=p_c=1/24$, and (c) $p=0.2 > p_c$. Symbols represents the numerical solutions ($\bigcirc$) of the rate equation in the non-steady-state limit and the Monte Carlo simulation results ($\textcolor{red}{\square}$) for given final system size $N=10^3$, respectively.
}
\label{fig:fig1}
\end{figure*}

\begin{figure}[!t]
\center
\includegraphics[width=1.0\linewidth]{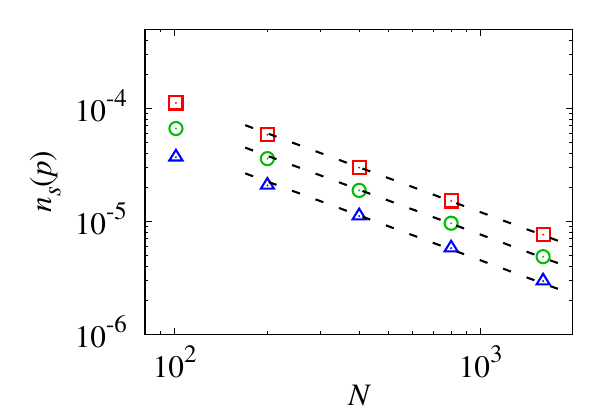}
\caption{(Color online)
Plots of the cluster size distribution $n_s(p)$ versus the final system size $N$ for $s=4~(\textcolor{red}{\square})$, $s=6~(\textcolor{green}{\bigcirc})$, and $s=8~(\textcolor{blue}{\triangle})$. Dotted lines are guidelines for power-law behavior. It confirms that clusters of even sizes vanish in the steady-state limit.
}
\label{fig:fig2}
\end{figure}

A model for a GRSC starts at $N_0$ isolated nodes and a new node is added at each time step $t$ where the total number of nodes is $N(t)=N_{0}+t$. Three nodes are randomly selected and then they are all connected with probability $p$ at each time step $t$. We consider this triangle with three selected nodes as $2$-simplex. The cluster size distribution $n_s(p,t)$ is defined as the number of clusters of size s divided by $N(t)$. The rate equation of $n_s(p,t)$ thus becomes
%
\begin{align}
\label{eq:rate_eq_of_n_s_p_t}
\frac{d (N(t)n_s(p,t))}{d t} =& p\Biggl[ \sum_{i,j,k=1}^{\infty}{in_i jn_j kn_k \delta_{i+j+k,s}} \nonumber \\
& + \sum_{i,j=1}^{\infty}{3in_i\frac{i}{N}jn_j\delta_{i+j,s}}- 3sn_s - 3sn_s\frac{s}{N}  \Biggr] \nonumber \\
&+\delta_{1s}.
\end{align}
The first and third terms of Eq.~\eqref{eq:rate_eq_of_n_s_p_t} correspond to the all three nodes belonging to three distinct clusters, respectively. It contributes that cluster size increases by an even-numbered size when the merging process occurs. Hence, when the system starts at all isolated nodes, the odd sizes of clusters only can exist from these terms. The second and fourth terms of Eq.~\eqref{eq:rate_eq_of_n_s_p_t}, however, correspond that two nodes belong to the same cluster among three selected nodes. It represents that cluster size can increase by size $1$ and then allow that there can exist clusters of all sizes for $s \ge 4$. The fifth term represents that a new node is added at each time step $t$. Numerical solutions of Eq.~\eqref{eq:rate_eq_of_n_s_p_t} are consistent with the Monte Carlo simulation results for the given final system size $N=10^3$ as shown in Fig.~\ref{fig:fig1}. To check the value of $n_s$ in the steady-state limit, we plotted $n_s$ versus $N(t)$ for $s=4, 6,$ and $8$ and confirm they clearly exhibit  power law behavior as shown in Fig.~\ref{fig:fig2}. It confirms that clusters of even sizes vanish in the steady-state limit.

\begin{figure*}[!t]
\center
\includegraphics[width=1.0\linewidth]{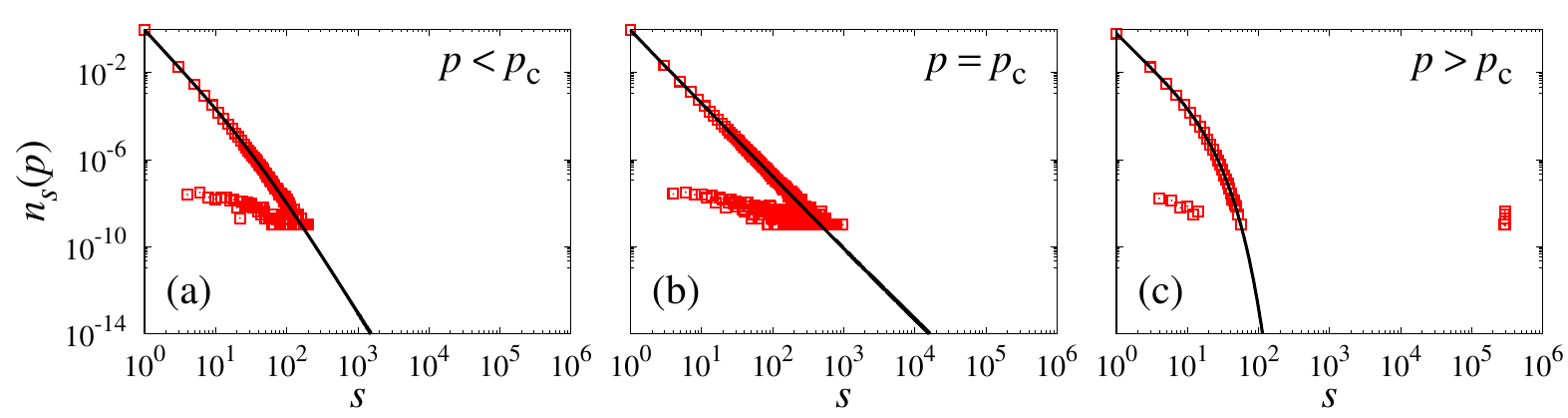}
\caption{(Color online)
Plots of the cluster size distribution $n_s(p)$ for (a) $p = 0.03 < p_c$, (b) $p=p_c=1/24$, and (c) $p=0.2 > p_c$. The black solid lines and red open square symbols ($\textcolor{red}{\square}$) represents the numerical solutions of the rate equations in the steady-state limit for the truncated size $s^{*} = 10^6$ and the Monte Carlo simulation results for given final system size $N=10^6$, respectively.
}
\label{fig:fig3}
\end{figure*}

\begin{figure}[!t]
\center
\includegraphics[width=1.0\linewidth]{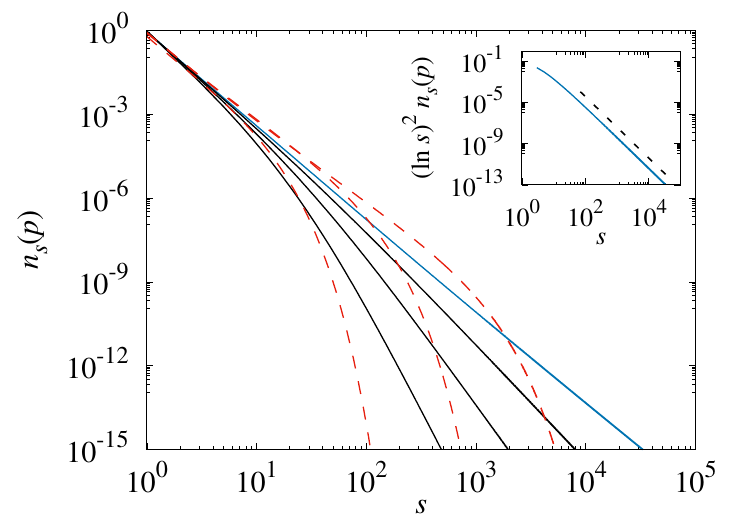}
\caption{(Color online)
Plot of $n_s (p)$ versus $s$ at $p = p_c = 1/24$ (blue solid line), for $p > p_c$ (red dashed curves), and for $p < p_c$ (black solid lines) based on numerical values obtained from the rate equation. For $p \le p_c$, $n_s (p)$ decays in a power-law manner, indicating that the transition is infinite order. Inset: Plot of $(\ln s)^2 n_s (p)$ versus $s$ at $p = p_c$. The black dashed line is guideline with slope $-3$.
}
\label{fig:fig4}
\end{figure}

Assuming that there are steady-state solutions of $n_s(p,t)$ as time $t$ goes to infinity, the cluster size distribution $n_s(p,t)$ is reduced to $n_s(p)$ independent of the time $t$. Moreover, the second and fourth terms in Eq \eqref{eq:rate_eq_of_n_s_p_t} can be neglected because they are intensive, and then there are non-zero solutions only for the odd sizes of clusters. The above Eq.~\eqref{eq:rate_eq_of_n_s_p_t} thus is reduced to the time-independent form as follows:
\begin{align}
\label{eq:rate_eq_of_n_s_p_t_in_steady_state}
n_s(p) &= p\Biggl[ \sum_{i,j,k=1}^{\infty}{in_i jn_j kn_k \delta_{i+j+k,s} } - 3sn_s \Biggr] +\delta_{1s},
\end{align}
Rearranging for $n_s(p)$, one can get
\begin{align}
\label{eq:sol_of_rate_eq_of_n_s_p_t_in_steady_state}
n_s(p) &= \frac{p\Bigl[ \sum_{i,j,k=1}^{\infty}{in_i jn_j kn_k \delta_{i+j+k,s}} \Bigr] +\delta_{1s}}{1+3sp},
\end{align}
Eq.~\eqref{eq:sol_of_rate_eq_of_n_s_p_t_in_steady_state} can be solved numerically up to the truncated size $s^{*}$, and the corresponding results for $s^{*}=10^6$ are consistent with Monte Carlo simulation results with the odd sizes of clusters for the given final system size $N=10^6$ as shown in Fig.~\ref{fig:fig3}. Moreover, we confirm there is critical phase where the cluster size distribution $n_s$ decays in the power-law manner for $p \le p_c$ and the estimated critical exponent $\tau$ at the critical point $p_c$ is $-3$ following the scaling forms of $n_s(p) \sim s^{-\tau} ln^{-2}s$ as shown in Fig.~\ref{fig:fig4}.

Now, we define the generating function for the probability $sn_s$, that a randomly chosen node belongs to the cluster of size s, defined as
\begin{align}
\label{eq:generating_function}
f(x) = \sum_{s=1}^{\infty} {sn_s x^s}.
\end{align}
The giant cluster $G$ and the second-order moment $\langle s \rangle$ are obtained as $G=1-\sum_{s}{sn_s}=1-f(1)$ and  $\langle s \rangle = \sum_{s}{s^2n_s} = f'(1)$, respectively. 

Multiplying both sides of Eq.~\eqref{eq:rate_eq_of_n_s_p_t_in_steady_state} by $s x^s$ and
sum over $s$, one can get
\begin{align}
\label{eq:rate_eq_with_generating_func}
f(x)=x-3pxf'(x) + 3pxf'(x)f^2(x),
\end{align}
where $f'=df/dx$. In another form, Eq.~\eqref{eq:rate_eq_with_generating_func} becomes
\begin{align}
\label{eq:rate_eq_with_generating_func_in_another_form}
f'(x)=\frac{1-\frac{f(x)}{x}} {3p(1-f^2(x))},
\end{align}

When $x=1$, the form of Eq.~\eqref{eq:rate_eq_with_generating_func_in_another_form} is classified into those in the normal and percolation phase. In the non-percolating phase with $f(1)=1$ yielding $G=1-f(1)=0$, Eq.~\eqref{eq:rate_eq_with_generating_func_in_another_form} becomes
\begin{align}
\label{eq:rate_eq_with_generating_func_in_another_form_when_x_1_in_normal_phase}
f'(1)=\frac{f'(1)-1} {6pf'(1)}.
\end{align}
The solution of Eq.~\eqref{eq:rate_eq_with_generating_func_in_another_form_when_x_1_in_normal_phase} is
\begin{align}
\label{eq:solution_of_rate_eq_with_generating_func_in_another_form_when_x_1_in_normal_phase}
f'(1)=\frac{1 \pm \sqrt{1-24p}} {12p}.
\end{align}
The solutions are valid only for $0 \le p \le 1/24$. Moreover, only the single solution with negative sign is valid since there must be only isolated nodes of size $1$ at $p=0$. Thus, $G=1-f(1)=0$ and $\langle s \rangle = f'(1)={(1 - \sqrt{1-24p})} / {12p}$ in the non-percolating phase for $0 \le p \le 1/24$. 

In the percolation phase with $f(1) \ne 1$ yielding $G=1-f(1)=1-\int_{0}^{1}{dx {f'(x)}}$, Eq \eqref{eq:rate_eq_with_generating_func_in_another_form} becomes 
\begin{align}
\label{eq:rate_eq_with_generating_func_in_another_form_when_x_1_in_percolation_phase}
f'(1)=\frac{1} {3p(1+f(1))},
\end{align}
where $f(1) = \int_{0}^{1}{dx {f'(x)}} = \int_{0}^{1}{dx {\frac{1-\frac{f(x)}{x}} {3p(1-f^2(x))}}}$. 

\begin{figure*}[!t]
\center
\includegraphics[width=1.0\linewidth]{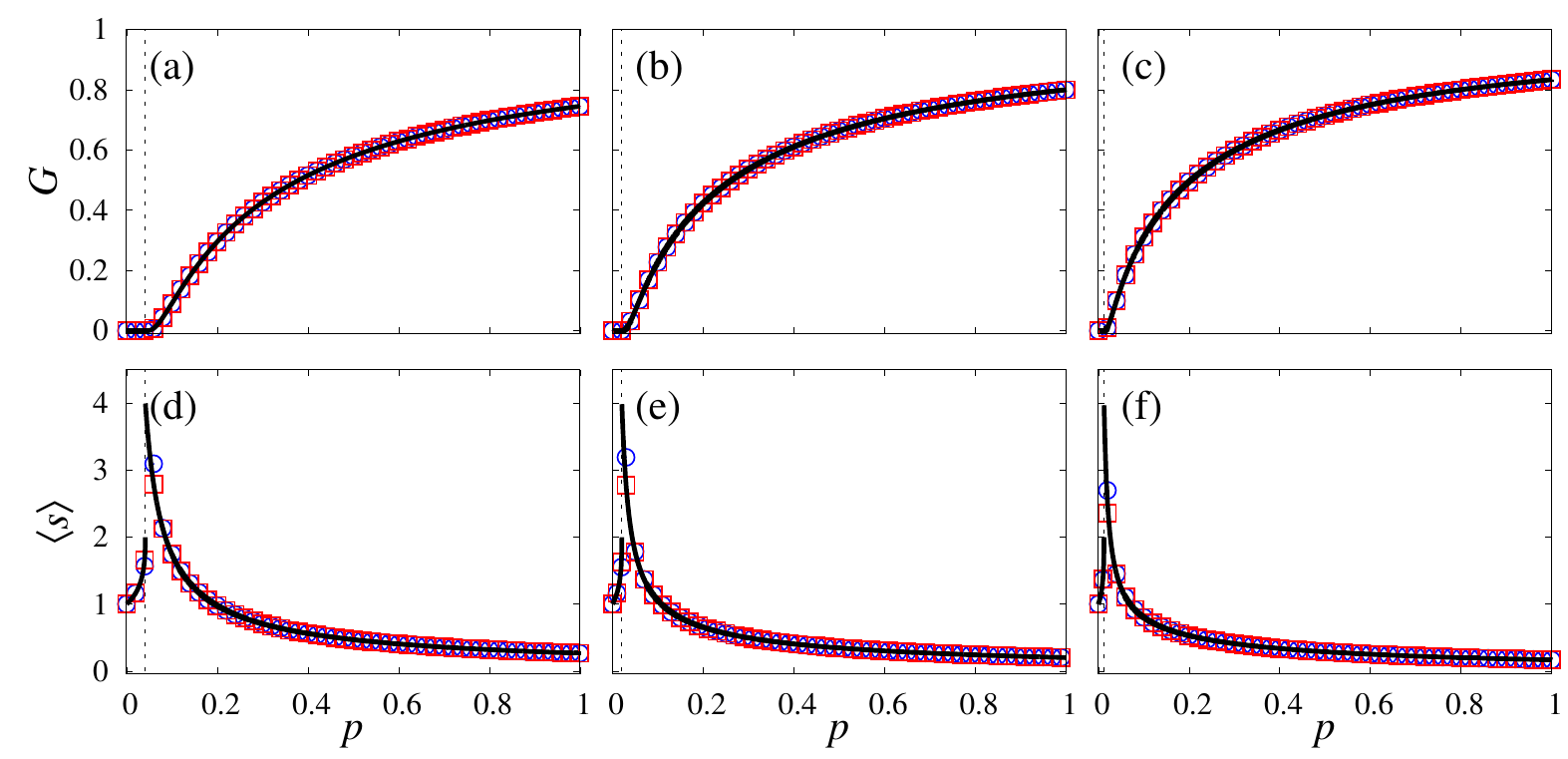}
\caption{(Color online)
Plots of the giant cluster size $G$ and the second-order moment $\langle s \rangle$ versus $p$ for $d=2$ in (a) and (d), $d=3$ in (b) and (e), and $d=4$ in (c) and (f). 
The blue open circle ($\textcolor{blue}{\bigcirc}$) and red open square symbols ($\textcolor{red}{\square}$) represent the Monte Carlo simulation results for the final system size $N=10^6$ and the numerical solutions of the rate equations in the steady-state limit, respectively. Black solid curves are calculated from $f(1)$ and $f'(1)$ for $G$ and $\langle s \rangle$. The transition point $p_c=1/(4d(d+1))$ is indicated by the dotted line for each given $d$.
}
\label{fig:fig5}
\end{figure*}

We thus define the transition point $p_c$ where the percolation arises at $p_c \equiv 1/24$. Finally, the forms of $G$ and $\langle s \rangle$ are described as follows:
\begin{align}
\label{eq:form_of_G}
G =
\begin{cases}
0  & \textmd{for}~~~~ p < p_c , \\
1 - \int_{0}^{1}{dx {\frac{1-\frac{f(x)}{x}} {3p(1-f^2(x))}}} & \textmd{for}~~~~ p \geq p_c ,
\end{cases}
\end{align}
\begin{align}
\label{eq:Second_moment_form_of_G}
\langle s \rangle =
\begin{cases}
\frac{1 - \sqrt{1-24p}} {12p}  & \textmd{for}~~~~ p < p_c , \\
\frac{1} {3p \Bigl(1 + \int_{0}^{1}{dx {\frac{1-\frac{f(x)}{x}} {3p(1-f^2(x))}}}  \Bigr)} & \textmd{for}~~~~ p \geq p_c ,
\end{cases}
\end{align}
where the transition point $p_c$ is defined as $p_c = 1/24$. The right-hand limit of ${\langle s \rangle}$ is $2$ and the right-hand limit of ${\langle s \rangle}$ becomes $4$ since $f(1) = 1$ at $p = p_c$. It shows clearly there is discontinuity in Eq.~\eqref{eq:Second_moment_form_of_G} at $p=p_c$. These solutions are consistent with those obtained by solving numerically the rate equation \eqref{eq:rate_eq_of_n_s_p_t_in_steady_state} and the Monte Carlo simulation results as shown in Figs. \ref{fig:fig5} (a), (d).

\subsection{Graph and facet degree distributions}
\label{subsec:Graph and facet degree distributions with 2-simplexes}

 The graph degree distribution $P_g(k,p,t)$ is defined as the number of nodes with degree $k$ divided by $N(t)$. The rate equation of $P_g(k,p,t)$ thus becomes

\begin{align}
\label{eq:rate_eq_of_P_k_p_t_1}
\frac{d (N(t)P_g(k,p,t))}{d t} &= -3p P_g(k) + \delta_{0k} & \textmd{for} ~~k < 2, \\
\label{eq:rate_eq_of_P_k_p_t_2}
\frac{d (N(t)P_g(k,p,t))}{d t} &= 3p(P_g(k-2) - P_g(k)) & \textmd{for} ~~k \ge 2.
\end{align}
This equation is valid for even $k$ including zero since degree always increases by the number $2$. Assuming $P_g(k,p,t)$ converges to constant value as time goes to infinity, the Eqs. \eqref{eq:rate_eq_of_P_k_p_t_1} and \eqref{eq:rate_eq_of_P_k_p_t_2} become
\begin{align}
\label{eq:rate_eq_of_P_k_p_t_in_steady_state}
P_g(k,p) &= -3p P_g(k) + \delta_{0k} & \textmd{for} ~~k < 2, \\
P_g(k,p) &= 3p(P_g(k-2) - P_g(k)) & \textmd{for} ~~k \ge 2.
\end{align}
One thus get $P_g(0,p)= \frac{1}{1+3p}$, $P_g(1,p)=0$, and $P_g(k,p)= \frac{3pP_g(k-2,p)}{1+3p}$ for $k \ge 2$ with leading to follow equations.
\begin{align}
\label{eq:solution_of_rate_eq_of_P_k_p_t_1}
P_g(k,p) &= 0 & \textmd{for odd} ~~k , \\
\label{eq:solution_of_rate_eq_of_P_k_p_t_2}
P_g(k,p) &= \frac{(3p)^{k/2}}{(1+3p)^{k/2 + 1}} & \textmd{for even} ~~k .
\end{align}

\begin{figure*}[!t]
\center
\includegraphics[width=0.8\linewidth]{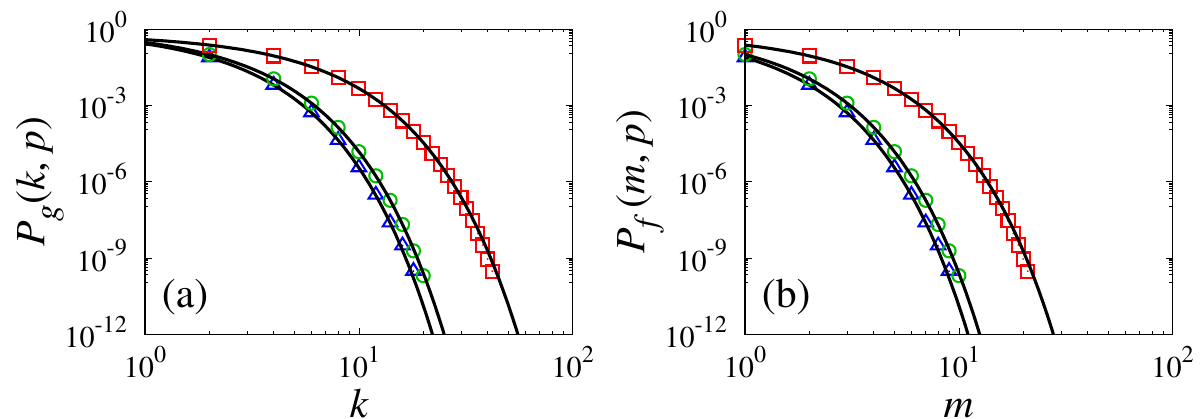}
\caption{(Color online)
Plots of (a) the degree distribution $P_g(k,p)$ versus $k$ and (b) the facet degree distribution $P_f(m,p)$ versus $m$. Symbols represent the Monte Carlo simulation results for the final system size $N=10^6$ at given $p = 0.03 ~(\textcolor{blue}{\triangle})$, $p=p_c=1/24 ~(\textcolor{green}{\bigcirc})$, and $p=0.2 ~(\textcolor{red}{\square})$, respectively. Solid curves are calculated from the solutions of the rate equations of the graph and facet degree distributions.
}
\label{fig:fig6}
\end{figure*}

Now, let's define the facet degree as the number of hyperedge connected to node. Then the facet degree distribution $P_f(m,p,t)$ is defined as the number of nodes with facet degree $m$ divided by $N(t)$. The rate equation of $P_f(m,p,t)$ thus becomes
\begin{align}
\label{eq:rate_eq_of_facet_P_m_p_t_1}
\frac{d (N(t)P_f(m,p,t))}{d t} &= -3p P_f(m) + \delta_{0m} & \textmd{for} ~~m < 1, \\
\label{eq:rate_eq_of_facet_P_m_p_t_2}
\frac{d (N(t)P_f(m,p,t))}{d t} &= 3p(P_f(m-1) - P_f(m)) & \textmd{for} ~~m \ge 1.
\end{align}
Assuming $P_f(m,p,t)$ converges to constant value as time goes to infinity, the Eqs. \eqref{eq:rate_eq_of_facet_P_m_p_t_1} and \eqref{eq:rate_eq_of_facet_P_m_p_t_2} become
\begin{align}
\label{eq:rate_eq_of_facet_P_m_p_t_in_steady_state}
P_f(m,p) &= -3p P_f(m) + \delta_{0m} & \textmd{for} ~~m < 1, \\
P_f(m,p) &= 3p(P_f(m-1) - P_f(m)) & \textmd{for} ~~m \ge 1.
\end{align}
One thus get $P_f(0,p)= \frac{1}{1+3p}$, and $P_f(m,p)= \frac{3pP_f(m-1,p)}{1+3p}$ for $m \ge 1$ with leading to follow equation.
\begin{align}
\label{eq:solution_of_rate_eq_of_facet_P_m_p_t}
P_f(m,p) &= \frac{(3p)^{m}}{(1+3p)^{m + 1}} .
\end{align}
Eqs. \eqref{eq:solution_of_rate_eq_of_P_k_p_t_1} and \eqref{eq:solution_of_rate_eq_of_P_k_p_t_2} for the degree and Eq.~\eqref{eq:solution_of_rate_eq_of_facet_P_m_p_t} for the face degree are consistent with the Monte Carlo simulation results for the final system size $N=10^6$ as shown in Fig.~\ref{fig:fig6}.

%

\section{GRSC model with $d$-simplexes}
\label{sec:GRSC model with $d$-simplexes}

\subsection{Models and rate equations}
\label{subsec:Models and rate equations with d-simplexes}

We consider the GSRC model where a $d$-simplex is established among $d+1$ nodes at each time step. The rate equation of $n_s(p,t)$ becomes 
\begin{align}
\label{eq:general_rate_eq_of_n_s_p_t}
&\frac{d (N(t)n_s(p,t))}{d t} \nonumber \\
=& p\Biggl[ \sum_{i_{1}, \ldots ,i_{d+1}=1}^{\infty}{\Bigl( \Pi_{\alpha=1}^{d+1} {i_{\alpha}n_{i_{\alpha}}} \Bigr) \delta_{\sum_{\alpha=1}^{d+1}{i_{\alpha}}, s}} \nonumber \\
&+ \sum_{r=1}^{d-1}  \sum_{ i_{1}, i_{r+2}, \ldots, i_{d+1}=1 }^{\infty}  \binom{d+1}{r+1} i_{1}n_{i_{1}} \Bigl(\frac{i_{1}}{N} \Bigr)^{r} \Bigl( \Pi_{\alpha=r+2}^{d+1} {i_{\alpha}n_{i_{\alpha}}} \Bigr)  \nonumber \\
&\times \delta_{i_{1} + \sum_{\alpha=r+2}^{d+1}{i_{\alpha}}, s}  -(d+1)sn_s -\sum_{r=1}^{d-1} {\binom{d+1}{r} sn_s \Bigl(\frac{s}{N}\Bigr)^r} \Biggr] \nonumber \\
&+\delta_{1s}.
\end{align}
The first and third terms of Eq.~\eqref{eq:general_rate_eq_of_n_s_p_t} represent that the selected $d+1$ nodes belong to $d+1$ distinct clusters. The second and fourth terms of Eq.~\eqref{eq:general_rate_eq_of_n_s_p_t} allow to select nodes in the same cluster. This model is reduced to the Callaway's growing random network model when $d=1$.

In the steady-state limit, time $t$ goes to infinity and then second and fourth terms of Eq.~\eqref{eq:general_rate_eq_of_n_s_p_t} become negligible. The rate equation thus can be written as follows:
\begin{align}
\label{eq:general_rate_eq_of_n_s_p_t_in_steady_state}
n_s(p) =& p\Biggl[ \sum_{i_{1}, \ldots ,i_{d+1}=1}^{\infty}{\Bigl( \Pi_{\alpha=1}^{d+1} {i_{\alpha}n_{i_{\alpha}}} \Bigr) \delta_{\sum_{\alpha=1}^{d+1}{i_{\alpha}}, s}} -(d+1)sn_s  \Biggr] \nonumber \\
&+\delta_{1s}.
\end{align}
Rearranging for $n_s(p)$, one can get 
\begin{align}
\label{eq:sol_of_general_rate_eq_of_n_s_p_t_in_steady_state}
n_s(p) &= \frac{p\Bigl[ \sum_{i_{1}, \ldots ,i_{d+1}=1}^{\infty}{\Bigl( \Pi_{\alpha=1}^{d+1} {i_{\alpha}n_{i_{\alpha}}} \Bigr) \delta_{\sum_{\alpha=1}^{d+1}{i_{\alpha}}, s}} \Bigr] +\delta_{1s}}{1-(d+1)sp}.
\end{align}
To use generating functions, multiplying both sides of Eq.~\eqref{eq:general_rate_eq_of_n_s_p_t_in_steady_state} by $s x^s$ and
sum over $s$, one can get
\begin{align}
\label{eq:general_rate_eq_with_generating_func}
f(x)=x-(d+1)pxf'(x) + (d+1)pxf'(x)f^{d+1}(x),
\end{align}
where $f'=df/dx$. In another form, Eq.~\eqref{eq:general_rate_eq_with_generating_func} becomes
\begin{align}
\label{eq:general_rate_eq_with_generating_func_in_another_form}
f'(x)=\frac{1-\frac{f(x)}{x}} {dp(1-f^{d}(x))},
\end{align}

When $x=1$, the form of Eq.~\eqref{eq:general_rate_eq_with_generating_func_in_another_form} is classified into those in the normal and percolation phase. In the non-percolating phase with $f(1)=1$ yielding $G=1-f(1)=0$, Eq.~\eqref{eq:general_rate_eq_with_generating_func_in_another_form} becomes
\begin{align}
\label{eq:general_rate_eq_with_generating_func_in_another_form_when_x_1_in_normal_phase}
f'(1)=\frac{f'(1)-1} {d(d+1)pf'(1)}.
\end{align}
The solution of Eq.~\eqref{eq:general_rate_eq_with_generating_func_in_another_form_when_x_1_in_normal_phase} is
\begin{align}
\label{eq:solution_of_general_rate_eq_with_generating_func_in_another_form_when_x_1_in_normal_phase}
f'(1)=\frac{1 \pm \sqrt{1-4d(d+1)p}} {2d(d+1)p}.
\end{align}
The solutions are valid only for $0 \le p \le 1/(4d(d+1))$. Moreover, only the single solution with negative sign is valid since there must be only isolated nodes of size $1$ at $p=0$. Thus, $G=1-f(1)=0$ and $\langle s \rangle = f'(1)={(1 - \sqrt{1-4d(d+1)p})} / {(2d(d+1)p)}$ in the non-percolating phase for $0 \le p \le 1/(4d(d+1))$. 

In the percolation phase with $f(1) \ne 1$ yielding $G=1-f(1)=1-\int_{0}^{1}{dx {f'(x)}}$, Eq \eqref{eq:general_rate_eq_with_generating_func_in_another_form} becomes 
\begin{align}
\label{eq:general_rate_eq_with_generating_func_in_another_form_when_x_1_in_percolation_phase}
f'(1)=\frac{1} {dp \sum_{i=0}^{d-1} {f^{i}(1)}},
\end{align}
where $f(1) = \int_{0}^{1}{dx {f'(x)}} = \int_{0}^{1}{dx {\frac{1-\frac{f(x)}{x}} {dp(1-f^{d}(x))}}}$. 

We thus define the transition point $p_c$ where the percolation arises as $p_c \equiv 1/(4d(d+1))$. Finally, the forms of $G$ and $\langle s \rangle$ are described as follows:
\begin{align}
\label{eq:general_form_of_G}
G =
\begin{cases}
0  & \textmd{for}~~~~ p < p_c , \\
1 - \int_{0}^{1}{dx {\frac{1-\frac{f(x)}{x}} {dp(1-f^{d}(x))}}} & \textmd{for}~~~~ p \geq p_c ,
\end{cases}
\end{align}
\begin{align}
\label{eq:general_Second_moment_form_of_G}
\langle s \rangle =
\begin{cases}
\frac{1 - \sqrt{1-4d(d+1)p}} {2d(d+1)p}.  & \textmd{for}~~~~ p < p_c , \\
f'(1)=\frac{1} {dp \sum_{i=0}^{d-1} { \Bigl( \int_{0}^{1}{dx {\frac{1-\frac{f(x)}{x}} {dp(1-f^{d}(x))}}} \Bigr) }^{i}} & \textmd{for}~~~~ p \geq p_c ,
\end{cases}
\end{align}
where the transition point $p_c$ is defined as $p_c = 1/(4d(d+1))$. The left-hand limit of ${\langle s \rangle}$ is $2$ and the right-hand limit of ${\langle s \rangle}$ becomes $4$ since $f(1)$ is always unity at $p = p_c$ regardless of $d$ values.   It shows clearly there is discontinuity in Eq.~\eqref{eq:general_Second_moment_form_of_G} at $p=p_c$. Eqs. \eqref{eq:general_form_of_G} and \eqref{eq:general_Second_moment_form_of_G} for $d=2, 3,$ and $4$ are consistent with those obtained by solving numerically the rate equation \eqref{eq:general_rate_eq_of_n_s_p_t_in_steady_state} and the Monte Carlo simulation results as shown in Fig.~\ref{fig:fig5}.

\subsection{Graph and facet degree distributions}
\label{subsec:Graph and facet degree distributions with d-simplexes}

The graph degree distribution $P_g(k,p,t)$ is defined as the number of nodes with degree $k$ divided by $N(t)$. The rate equations of $P_g(k,p,t)$ are written as
\begin{align}
\label{eq:general_rate_eq_of_P_k_p_t_1}
\frac{d (N(t)P_g(k,p,t))}{d t} =& -(d+1)p P_g(k) + \delta_{0k} \nonumber \\
&\hspace{2.5cm} \textmd{for} ~~k < d, \\
\label{eq:general_rate_eq_of_P_k_p_t_2}
\frac{d (N(t)P_g(k,p,t))}{d t} =& (d+1)p(P_g(k-d) - P_g(k)) \nonumber \\
&\hspace{2.5cm} \textmd{for} ~~k \ge d.
\end{align}
This equation is valid for $k = dn$ with a non-negative integer $n \ge 0$ including zero since degree always increases by the number $d$. Assuming $P_g(k,p,t)$ converges to constant value of $P_g(k,p)$ as time goes to infinity, The Eqs. \eqref{eq:general_rate_eq_of_P_k_p_t_1} and \eqref{eq:general_rate_eq_of_P_k_p_t_2} become
\begin{align}
\label{eq:general_rate_eq_of_P_k_p_t_in_steady_state}
P_g(k,p) &= -(d+1)p P_g(k) + \delta_{0k} & \textmd{for} ~~k < d, \\
P_g(k,p) &= (d+1)p(P_g(k-d) - P_g(k)) & \textmd{for} ~~k \ge d.
\end{align}
One thus get $P_g(0,p)= \frac{1}{1+(d+1)p}$, $P_g(1,p)=0$, \ldots, $P_g(d-1,p)=0$ and $P_g(k,p)= \frac{(d+1)pP_g(k-d, p)}{1+(d+1)p}$ for $k \ge d$ with leading to follow equations.
\begin{align}
\label{eq:solution_of_general_rate_eq_of_P_k_p_t_1}
P_g(k,p) &= 0 & \textmd{for} ~~ k \ne dn , \\
\label{eq:solution_of_general_rate_eq_of_P_k_p_t_2}
P_g(k,p) &= \frac{((d+1)p)^{k/d}}{(1+(d+1)p)^{k/d + 1}} & \textmd{for} ~~ k = dn ,
\end{align}
where $n$ is a non-negative integer including zero.

Now, let's define the facet degree as the number of hyperedge connected to node. Then the facet degree distribution $P_f(m,p,t)$ is defined as the number of nodes with facet degree $m$ divided by $N(t)$. The rate equation of $P_f(m,p,t)$ thus becomes
\begin{align}
\label{eq:general_rate_eq_of_facet_P_m_p_t_1}
\frac{d (N(t)P_f(m,p,t))}{d t} =& -(d+1)p P_f(m) + \delta_{0m} \nonumber \\
&\hspace{2.5cm} \textmd{for} ~~m < 1, \\
\label{eq:general_rate_eq_of_facet_P_m_p_t_2}
\frac{d (N(t)P_f(m,p,t))}{d t} =& (d+1)p(P_f(m-1) - P_f(m)) \nonumber \\
&\hspace{2.5cm} \textmd{for} ~~m \ge 1. 
\end{align}
Assuming $P_f(m,p,t)$ converges to constant value of $P_f(m,p)$ as time goes to infinity, The Eqs. \eqref{eq:general_rate_eq_of_facet_P_m_p_t_1} and \eqref{eq:general_rate_eq_of_facet_P_m_p_t_2} become
\begin{align}
\label{eq:general_rate_eq_of_facet_P_m_p_t_in_steady_state}
P_f(m,p) &= -(d+1)p P_f(m) + \delta_{0m} & \textmd{for} ~~m < 1, \\
P_f(m,p) &= (d+1)p(P_f(m-1) - P_f(m)) & \textmd{for} ~~m \ge 1.
\end{align}
One thus get $P_f(0,p)= \frac{1}{1+(d+1)p}$, and $P_f(m,p)= \frac{(d+1)pP_f(m-1,p)}{1+(d+1)p}$ for $m \ge 1$ with leading to follow equation.
\begin{align}
\label{eq:solution_of_general_rate_eq_of_facet_P_m_p_t}
P_f(m,p) &= \frac{((d+1)p)^{m}}{(1+(d+1)p)^{m + 1}} .
\end{align}
%

\subsection{Giant cluster size}
\label{subsec:Giant cluster size with d-simplexes}

We derive the explicit form of the giant cluster size in terms of $p$ and $d$. Eq.~\eqref{eq:general_rate_eq_with_generating_func_in_another_form} can be written in the form
\begin{align}
\label{eq:general_rate_eq_with_generating_func_in_terms_of_F_and_X}
F'(X)= \frac{1-(1-F)/(1-X)}{(d+1)p(1-(1-F)^{d})},
\end{align}
where $F(X) \equiv 1-f(x)$ and $X \equiv 1-x$. Near the critical point $p=p_{c}$ and $x=1$, Eq.~\eqref{eq:general_rate_eq_with_generating_func_in_terms_of_F_and_X} becomes
\begin{align}
\label{eq:general_rate_eq_with_generating_func_in_terms_of_F_and_X_around_pcand_x_1}
F(X)F'(X)= \frac{1}{d(d+1)p} (F-X),
\end{align}
where $F(0) = 1-f(1) = G$. The solution of Eq.~\eqref{eq:general_rate_eq_with_generating_func_in_terms_of_F_and_X_around_pcand_x_1} for $p > p_c=1/(4d(d+1))$ is
\begin{align}
\label{eq:solution_of_eq:general_rate_eq_with_generating_func_in_terms_of_F_and_X_around_pc_and_x_1}
&\frac{-1}{\sqrt{4d(d+1)p-1}}\arctan{\Bigl[ \frac{2d(d+1)p(F/X) - 1}{\sqrt{4d(d+1)p-1}} \Bigr]} \nonumber \\
&- \ln{\sqrt{X^{2}-FX+d(d+1)pF^{2}}} = C ,
\end{align}
where $C$ is an integral constant. Around $X=0$, one can get 
\begin{align}
\label{eq:solution2_of_eq:general_rate_eq_with_generating_func_in_terms_of_F_and_X_around_pc_and_x_1}
\frac{-\pi/2}{\sqrt{4d(d+1)p-1}} - \ln{\sqrt{d(d+1)p}} - \ln{G} = C.
\end{align}
Moreover, from the form of Eq.~\eqref{eq:solution_of_eq:general_rate_eq_with_generating_func_in_terms_of_F_and_X_around_pc_and_x_1} in the limit of $X \gg G$ for $F < 2X$, one get
\begin{align}
\label{eq:solution3_of_eq:general_rate_eq_with_generating_func_in_terms_of_F_and_X_around_pc_and_x_1}
&\frac{-\pi/2}{\sqrt{4d(d+1)p-1}} - \ln{\sqrt{d(d+1)p}} - \ln{G} \nonumber \\
&= \frac{\pi/2}{\sqrt{4d(d+1)p-1}} - \frac{1}{1-F/(2X)} - \ln{X-\frac{F}{2}},
\end{align}
using the relation of $\arctan(x) = -(\pi/2) - \arctan(1/x)$ for $x<0$. 

Now, we also describe the threshold solution of Eq.~\eqref{eq:general_rate_eq_with_generating_func_in_terms_of_F_and_X_around_pcand_x_1} as follows
\begin{align}
\label{eq:Thgeneral_rate_eq_with_generating_func_in_other_form_around_pc}
F(x) = 2X(1-g(X)),
\end{align}
where $g(X)$ satisfies $\ln(Xg) + 1/g = \ln(c)$ and $c$ is a constant. Thus, around $p=p_c=1/(4d(d+1))$, Eq.~\eqref{eq:solution3_of_eq:general_rate_eq_with_generating_func_in_terms_of_F_and_X_around_pc_and_x_1} become
\begin{align}
\label{eq:solution4_of_eq:general_rate_eq_with_generating_func_in_terms_of_F_and_X_around_pc_and_x_1}
\frac{-\pi}{\sqrt{4d(d+1)p-1}} - \ln{\frac{G}{2}} = -\frac{1}{g} - \ln{(Xg)} = -\ln{c}.
\end{align}
Finally, one arrive at the explicit form of $G$ in terms of $p$ around the critical point as follows:
\begin{align}
\label{eq:final_solution_of_eq:general_rate_eq_with_generating_func_in_terms_of_F_and_X_around_pc_and_x_1}
G=2c\exp{\Bigl[ -\frac{\pi}{2\sqrt{d(d+1)}} ( p - p_c )^{-1/2} \Bigr]},
\end{align}
where the critical point $p_c$ is $p_c = 1/(4d(d+1))$.

\subsection{GRSC model with $(d+1)$-sided polygons}
\label{subsec:GRSC model with $(d+1)$-sided polygons}

When we consider a GRSC where a $(d+1)$-sided polygon is established among $d+1$ nodes to the system, all properties are the same as those of $d$-simplex but for the graph degree distribution. The graph degree distribution can be written as follows because only two links are added to each node when a polygon is established.
\begin{align}
\label{eq:solution_of_general_rate_eq_of_P_k_p_t_1_for_d_sided_plygon}
P_g(k,p) &= 0 & \textmd{for} ~~ k \ne 2n , \\
\label{eq:solution_of_general_rate_eq_of_P_k_p_t_2_for_d_sided_plygon}
P_g(k,p) &= \frac{((d+1)p)^{k/2}}{(1+(d+1)p)^{k/2 + 1}} & \textmd{for} ~~ k = 2n ,
\end{align}
where $n$ is a non-negative integer including zero.

\section{Conclusions}
\label{sec:Conclusions}

We proposed GRSC models capturing the fact that real-world systems are growing and include higher-order interactions. We first considered GRSCs with $2$-simplexes established with time, and then generalized them with $d$-simplexes for any positive integer $d$. Developing rate equations of cluster size distributions, we derived analytical forms of giant cluster sizes and second-order moments. Around percolation thresholds, the giant cluster emerges smoothly and the second-order moment is discontinuous with two finite distinct limit values. Subsequently, we obtained graph and facet degree distributions and confirmed that they are exponential. The percolation thresholds are more advanced with larger $d$ as higher-order interactions among nodes accelerate the growth of the system. All these properties guarantee that a GRSC exhibits an infinite-order phase transition regardless of the dimension of established simplexes. We strongly anticipate that our findings can lay the foundation for interpreting growing complex systems with higher-order interactions.

\vspace{-0.15cm}
\begin{acknowledgments}
\vspace{-0.2cm}
This work was supported by the National Research Foundation in Korea with Grant No. RS-2023-00279802 (B.K.), No. NRF-2021R1A6A3A01087148 (S.M.O.), and KENTECH Research Grant No. KRG2021-01-007 (B.K.) at Korea Institute of Energy Technology.
\end{acknowledgments}

\vspace{-0.15cm}
\section*{References}
\vspace{-0.75cm}
\bibliographystyle{apsrev4-2}
\bibliography{references.bib}

\begin{thebibliography}{39}%
\makeatletter
\providecommand \@ifxundefined [1]{%
 \@ifx{#1\undefined}
}%
\providecommand \@ifnum [1]{%
 \ifnum #1\expandafter \@firstoftwo
 \else \expandafter \@secondoftwo
 \fi
}%
\providecommand \@ifx [1]{%
 \ifx #1\expandafter \@firstoftwo
 \else \expandafter \@secondoftwo
 \fi
}%
\providecommand \natexlab [1]{#1}%
\providecommand \enquote  [1]{``#1''}%
\providecommand \bibnamefont  [1]{#1}%
\providecommand \bibfnamefont [1]{#1}%
\providecommand \citenamefont [1]{#1}%
\providecommand \href@noop [0]{\@secondoftwo}%
\providecommand \href [0]{\begingroup \@sanitize@url \@href}%
\providecommand \@href[1]{\@@startlink{#1}\@@href}%
\providecommand \@@href[1]{\endgroup#1\@@endlink}%
\providecommand \@sanitize@url [0]{\catcode `\\12\catcode `\$12\catcode `\&12\catcode `\#12\catcode `\^12\catcode `\_12\catcode `\%12\relax}%
\providecommand \@@startlink[1]{}%
\providecommand \@@endlink[0]{}%
\providecommand \url  [0]{\begingroup\@sanitize@url \@url }%
\providecommand \@url [1]{\endgroup\@href {#1}{\urlprefix }}%
\providecommand \urlprefix  [0]{URL }%
\providecommand \Eprint [0]{\href }%
\providecommand \doibase [0]{https://doi.org/}%
\providecommand \selectlanguage [0]{\@gobble}%
\providecommand \bibinfo  [0]{\@secondoftwo}%
\providecommand \bibfield  [0]{\@secondoftwo}%
\providecommand \translation [1]{[#1]}%
\providecommand \BibitemOpen [0]{}%
\providecommand \bibitemStop [0]{}%
\providecommand \bibitemNoStop [0]{.\EOS\space}%
\providecommand \EOS [0]{\spacefactor3000\relax}%
\providecommand \BibitemShut  [1]{\csname bibitem#1\endcsname}%
\let\auto@bib@innerbib\@empty
\bibitem [{\citenamefont {Barabási}\ and\ \citenamefont {Albert}(1999)}]{barabasi_emergence_1999}%
  \BibitemOpen
  \bibfield  {author} {\bibinfo {author} {\bibfnamefont {A.-L.}\ \bibnamefont {Barabási}}\ and\ \bibinfo {author} {\bibfnamefont {R.}~\bibnamefont {Albert}},\ }\href {https://doi.org/10.1126/science.286.5439.509} {\bibfield  {journal} {\bibinfo  {journal} {Science}\ }\textbf {\bibinfo {volume} {286}},\ \bibinfo {pages} {509} (\bibinfo {year} {1999})}\BibitemShut {NoStop}%
\bibitem [{\citenamefont {Newman}(2003)}]{newman_structure_2003}%
  \BibitemOpen
  \bibfield  {author} {\bibinfo {author} {\bibfnamefont {M.~E.~J.}\ \bibnamefont {Newman}},\ }\href {https://doi.org/10.1137/S003614450342480} {\bibfield  {journal} {\bibinfo  {journal} {SIAM Review}\ }\textbf {\bibinfo {volume} {45}},\ \bibinfo {pages} {167} (\bibinfo {year} {2003})}\BibitemShut {NoStop}%
\bibitem [{\citenamefont {Boccaletti}\ \emph {et~al.}(2006)\citenamefont {Boccaletti}, \citenamefont {Latora}, \citenamefont {Moreno}, \citenamefont {Chavez},\ and\ \citenamefont {Hwang}}]{boccaletti_complex_2006}%
  \BibitemOpen
  \bibfield  {author} {\bibinfo {author} {\bibfnamefont {S.}~\bibnamefont {Boccaletti}}, \bibinfo {author} {\bibfnamefont {V.}~\bibnamefont {Latora}}, \bibinfo {author} {\bibfnamefont {Y.}~\bibnamefont {Moreno}}, \bibinfo {author} {\bibfnamefont {M.}~\bibnamefont {Chavez}},\ and\ \bibinfo {author} {\bibfnamefont {D.~U.}\ \bibnamefont {Hwang}},\ }\href {https://doi.org/10.1016/j.physrep.2005.10.009} {\bibfield  {journal} {\bibinfo  {journal} {Physics Reports}\ }\textbf {\bibinfo {volume} {424}},\ \bibinfo {pages} {175} (\bibinfo {year} {2006})}\BibitemShut {NoStop}%
\bibitem [{\citenamefont {Granovetter}(1973)}]{granovetter_strength_1973}%
  \BibitemOpen
  \bibfield  {author} {\bibinfo {author} {\bibfnamefont {M.~S.}\ \bibnamefont {Granovetter}},\ }\href {https://doi.org/https://doi.org/10.1086/225469} {\bibfield  {journal} {\bibinfo  {journal} {American Journal of Sociology}\ }\textbf {\bibinfo {volume} {78}},\ \bibinfo {pages} {1360} (\bibinfo {year} {1973})}\BibitemShut {NoStop}%
\bibitem [{\citenamefont {Watts}\ and\ \citenamefont {Strogatz}(1998)}]{watts_collective_1998}%
  \BibitemOpen
  \bibfield  {author} {\bibinfo {author} {\bibfnamefont {D.~J.}\ \bibnamefont {Watts}}\ and\ \bibinfo {author} {\bibfnamefont {S.~H.}\ \bibnamefont {Strogatz}},\ }\href {https://doi.org/10.1038/30918} {\bibfield  {journal} {\bibinfo  {journal} {Nature}\ }\textbf {\bibinfo {volume} {393}},\ \bibinfo {pages} {440} (\bibinfo {year} {1998})}\BibitemShut {NoStop}%
\bibitem [{\citenamefont {Newman}\ and\ \citenamefont {Watts}(1999)}]{newman_scaling_1999}%
  \BibitemOpen
  \bibfield  {author} {\bibinfo {author} {\bibfnamefont {M.~E.~J.}\ \bibnamefont {Newman}}\ and\ \bibinfo {author} {\bibfnamefont {D.~J.}\ \bibnamefont {Watts}},\ }\href {https://doi.org/10.1103/PhysRevE.60.7332} {\bibfield  {journal} {\bibinfo  {journal} {Physical Review E}\ }\textbf {\bibinfo {volume} {60}},\ \bibinfo {pages} {7332} (\bibinfo {year} {1999})}\BibitemShut {NoStop}%
\bibitem [{\citenamefont {Moody}(2004)}]{moody_structure_2004}%
  \BibitemOpen
  \bibfield  {author} {\bibinfo {author} {\bibfnamefont {J.}~\bibnamefont {Moody}},\ }\href {https://doi.org/10.1177/000312240406900204} {\bibfield  {journal} {\bibinfo  {journal} {American Sociological Review}\ }\textbf {\bibinfo {volume} {69}},\ \bibinfo {pages} {213} (\bibinfo {year} {2004})}\BibitemShut {NoStop}%
\bibitem [{\citenamefont {Greening}\ \emph {et~al.}(2015)\citenamefont {Greening}, \citenamefont {Pinter-Wollman},\ and\ \citenamefont {Fefferman}}]{greening_higher-order_2015}%
  \BibitemOpen
  \bibfield  {author} {\bibinfo {author} {\bibfnamefont {B.~R.}\ \bibnamefont {Greening}, \bibfnamefont {Jr}}, \bibinfo {author} {\bibfnamefont {N.}~\bibnamefont {Pinter-Wollman}},\ and\ \bibinfo {author} {\bibfnamefont {N.~H.}\ \bibnamefont {Fefferman}},\ }\href {https://doi.org/10.1093/czoolo/61.1.114} {\bibfield  {journal} {\bibinfo  {journal} {Current Zoology}\ }\textbf {\bibinfo {volume} {61}},\ \bibinfo {pages} {114} (\bibinfo {year} {2015})}\BibitemShut {NoStop}%
\bibitem [{\citenamefont {Bianconi}(2021)}]{bianconi_higher-order_2021}%
  \BibitemOpen
  \bibfield  {author} {\bibinfo {author} {\bibfnamefont {G.}~\bibnamefont {Bianconi}},\ }\bibfield  {journal} {\bibinfo  {journal} {Elements in the Structure and Dynamics of Complex Networks}\ }\href {https://doi.org/10.1017/9781108770996} {10.1017/9781108770996} (\bibinfo {year} {2021})\BibitemShut {NoStop}%
\bibitem [{\citenamefont {Solé}\ \emph {et~al.}(2002)\citenamefont {Solé}, \citenamefont {Pastor-Satorras}, \citenamefont {Smith},\ and\ \citenamefont {Kepler}}]{sole_model_2002}%
  \BibitemOpen
  \bibfield  {author} {\bibinfo {author} {\bibfnamefont {R.~V.}\ \bibnamefont {Solé}}, \bibinfo {author} {\bibfnamefont {R.}~\bibnamefont {Pastor-Satorras}}, \bibinfo {author} {\bibfnamefont {E.}~\bibnamefont {Smith}},\ and\ \bibinfo {author} {\bibfnamefont {T.~B.}\ \bibnamefont {Kepler}},\ }\href {https://doi.org/10.1142/S021952590200047X} {\bibfield  {journal} {\bibinfo  {journal} {Advances in Complex Systems}\ }\textbf {\bibinfo {volume} {05}},\ \bibinfo {pages} {43} (\bibinfo {year} {2002})}\BibitemShut {NoStop}%
\bibitem [{\citenamefont {Vázquez}\ \emph {et~al.}(2002)\citenamefont {Vázquez}, \citenamefont {Flammini}, \citenamefont {Maritan},\ and\ \citenamefont {Vespignani}}]{vazquez_modeling_2002}%
  \BibitemOpen
  \bibfield  {author} {\bibinfo {author} {\bibfnamefont {A.}~\bibnamefont {Vázquez}}, \bibinfo {author} {\bibfnamefont {A.}~\bibnamefont {Flammini}}, \bibinfo {author} {\bibfnamefont {A.}~\bibnamefont {Maritan}},\ and\ \bibinfo {author} {\bibfnamefont {A.}~\bibnamefont {Vespignani}},\ }\href {https://doi.org/10.1159/000067642} {\bibfield  {journal} {\bibinfo  {journal} {Complexus}\ }\textbf {\bibinfo {volume} {1}},\ \bibinfo {pages} {38} (\bibinfo {year} {2002})}\BibitemShut {NoStop}%
\bibitem [{\citenamefont {Hartwell}\ \emph {et~al.}(1999)\citenamefont {Hartwell}, \citenamefont {Hopfield}, \citenamefont {Leibler},\ and\ \citenamefont {Murray}}]{hartwell_molecular_1999}%
  \BibitemOpen
  \bibfield  {author} {\bibinfo {author} {\bibfnamefont {L.~H.}\ \bibnamefont {Hartwell}}, \bibinfo {author} {\bibfnamefont {J.~J.}\ \bibnamefont {Hopfield}}, \bibinfo {author} {\bibfnamefont {S.}~\bibnamefont {Leibler}},\ and\ \bibinfo {author} {\bibfnamefont {A.~W.}\ \bibnamefont {Murray}},\ }\href {https://doi.org/10.1038/35011540} {\bibfield  {journal} {\bibinfo  {journal} {Nature}\ }\textbf {\bibinfo {volume} {402}},\ \bibinfo {pages} {C47} (\bibinfo {year} {1999})}\BibitemShut {NoStop}%
\bibitem [{\citenamefont {Spirin}\ and\ \citenamefont {Mirny}(2003)}]{spirin_protein_2003}%
  \BibitemOpen
  \bibfield  {author} {\bibinfo {author} {\bibfnamefont {V.}~\bibnamefont {Spirin}}\ and\ \bibinfo {author} {\bibfnamefont {L.~A.}\ \bibnamefont {Mirny}},\ }\href {https://doi.org/10.1073/pnas.2032324100} {\bibfield  {journal} {\bibinfo  {journal} {Proceedings of the National Academy of Sciences}\ }\textbf {\bibinfo {volume} {100}},\ \bibinfo {pages} {12123} (\bibinfo {year} {2003})}\BibitemShut {NoStop}%
\bibitem [{\citenamefont {Barabási}\ and\ \citenamefont {Oltvai}(2004)}]{barabasi_network_2004}%
  \BibitemOpen
  \bibfield  {author} {\bibinfo {author} {\bibfnamefont {A.-L.}\ \bibnamefont {Barabási}}\ and\ \bibinfo {author} {\bibfnamefont {Z.~N.}\ \bibnamefont {Oltvai}},\ }\href {https://doi.org/10.1038/nrg1272} {\bibfield  {journal} {\bibinfo  {journal} {Nature Reviews Genetics}\ }\textbf {\bibinfo {volume} {5}},\ \bibinfo {pages} {101} (\bibinfo {year} {2004})}\BibitemShut {NoStop}%
\bibitem [{\citenamefont {Goh}\ \emph {et~al.}(2005)\citenamefont {Goh}, \citenamefont {Kahng},\ and\ \citenamefont {Kim}}]{goh_graph_2005}%
  \BibitemOpen
  \bibfield  {author} {\bibinfo {author} {\bibfnamefont {K.~I.}\ \bibnamefont {Goh}}, \bibinfo {author} {\bibfnamefont {B.}~\bibnamefont {Kahng}},\ and\ \bibinfo {author} {\bibfnamefont {D.}~\bibnamefont {Kim}},\ }\href {https://doi.org/10.1016/j.physa.2005.03.044} {\bibfield  {journal} {\bibinfo  {journal} {Physica A: Statistical Mechanics and its Applications}\ }\textbf {\bibinfo {volume} {357}},\ \bibinfo {pages} {501} (\bibinfo {year} {2005})}\BibitemShut {NoStop}%
\bibitem [{\citenamefont {Kim}\ and\ \citenamefont {Kahng}(2010)}]{kim_fractal_2010}%
  \BibitemOpen
  \bibfield  {author} {\bibinfo {author} {\bibfnamefont {P.}~\bibnamefont {Kim}}\ and\ \bibinfo {author} {\bibfnamefont {B.}~\bibnamefont {Kahng}},\ }\href {https://doi.org/10.3938/jkps.56.1020} {\bibfield  {journal} {\bibinfo  {journal} {Journal of the Korean Physical Society}\ }\textbf {\bibinfo {volume} {56}},\ \bibinfo {pages} {1020} (\bibinfo {year} {2010})}\BibitemShut {NoStop}%
\bibitem [{\citenamefont {Ispolatov}\ \emph {et~al.}(2005)\citenamefont {Ispolatov}, \citenamefont {Krapivsky},\ and\ \citenamefont {Yuryev}}]{ispolatov_duplication-divergence_2005}%
  \BibitemOpen
  \bibfield  {author} {\bibinfo {author} {\bibfnamefont {I.}~\bibnamefont {Ispolatov}}, \bibinfo {author} {\bibfnamefont {P.~L.}\ \bibnamefont {Krapivsky}},\ and\ \bibinfo {author} {\bibfnamefont {A.}~\bibnamefont {Yuryev}},\ }\href {https://doi.org/10.1103/PhysRevE.71.061911} {\bibfield  {journal} {\bibinfo  {journal} {Physical Review E}\ }\textbf {\bibinfo {volume} {71}},\ \bibinfo {pages} {061911} (\bibinfo {year} {2005})}\BibitemShut {NoStop}%
\bibitem [{\citenamefont {Kim}\ \emph {et~al.}(2002)\citenamefont {Kim}, \citenamefont {Krapivsky}, \citenamefont {Kahng},\ and\ \citenamefont {Redner}}]{kim_infinite-order_2002}%
  \BibitemOpen
  \bibfield  {author} {\bibinfo {author} {\bibfnamefont {J.}~\bibnamefont {Kim}}, \bibinfo {author} {\bibfnamefont {P.~L.}\ \bibnamefont {Krapivsky}}, \bibinfo {author} {\bibfnamefont {B.}~\bibnamefont {Kahng}},\ and\ \bibinfo {author} {\bibfnamefont {S.}~\bibnamefont {Redner}},\ }\href {https://doi.org/10.1103/PhysRevE.66.055101} {\bibfield  {journal} {\bibinfo  {journal} {Physical Review E}\ }\textbf {\bibinfo {volume} {66}},\ \bibinfo {pages} {055101} (\bibinfo {year} {2002})}\BibitemShut {NoStop}%
\bibitem [{\citenamefont {Oh}\ \emph {et~al.}(2018)\citenamefont {Oh}, \citenamefont {Son},\ and\ \citenamefont {Kahng}}]{oh_suppression_2018}%
  \BibitemOpen
  \bibfield  {author} {\bibinfo {author} {\bibfnamefont {S.~M.}\ \bibnamefont {Oh}}, \bibinfo {author} {\bibfnamefont {S.-W.}\ \bibnamefont {Son}},\ and\ \bibinfo {author} {\bibfnamefont {B.}~\bibnamefont {Kahng}},\ }\href {https://doi.org/10.1103/PhysRevE.98.060301} {\bibfield  {journal} {\bibinfo  {journal} {Physical Review E}\ }\textbf {\bibinfo {volume} {98}},\ \bibinfo {pages} {060301} (\bibinfo {year} {2018})}\BibitemShut {NoStop}%
\bibitem [{\citenamefont {Oh}\ \emph {et~al.}(2019)\citenamefont {Oh}, \citenamefont {Son},\ and\ \citenamefont {Kahng}}]{oh_discontinuous_2019}%
  \BibitemOpen
  \bibfield  {author} {\bibinfo {author} {\bibfnamefont {S.~M.}\ \bibnamefont {Oh}}, \bibinfo {author} {\bibfnamefont {S.-W.}\ \bibnamefont {Son}},\ and\ \bibinfo {author} {\bibfnamefont {B.}~\bibnamefont {Kahng}},\ }\href {https://doi.org/10.1088/1742-5468/ab3110} {\bibfield  {journal} {\bibinfo  {journal} {Journal of Statistical Mechanics: Theory and Experiment}\ }\textbf {\bibinfo {volume} {2019}},\ \bibinfo {pages} {083502} (\bibinfo {year} {2019})}\BibitemShut {NoStop}%
\bibitem [{\citenamefont {Redner}(1998)}]{redner_how_1998}%
  \BibitemOpen
  \bibfield  {author} {\bibinfo {author} {\bibfnamefont {S.}~\bibnamefont {Redner}},\ }\href {https://doi.org/10.1007/s100510050359} {\bibfield  {journal} {\bibinfo  {journal} {The European Physical Journal B - Condensed Matter and Complex Systems}\ }\textbf {\bibinfo {volume} {4}},\ \bibinfo {pages} {131} (\bibinfo {year} {1998})}\BibitemShut {NoStop}%
\bibitem [{\citenamefont {Newman}(2001)}]{newman_structure_2001}%
  \BibitemOpen
  \bibfield  {author} {\bibinfo {author} {\bibfnamefont {M.~E.~J.}\ \bibnamefont {Newman}},\ }\href {https://doi.org/10.1073/pnas.98.2.404} {\bibfield  {journal} {\bibinfo  {journal} {Proceedings of the National Academy of Sciences}\ }\textbf {\bibinfo {volume} {98}},\ \bibinfo {pages} {404} (\bibinfo {year} {2001})}\BibitemShut {NoStop}%
\bibitem [{\citenamefont {Newman}(2004)}]{newman_coauthorship_2004}%
  \BibitemOpen
  \bibfield  {author} {\bibinfo {author} {\bibfnamefont {M.~E.~J.}\ \bibnamefont {Newman}},\ }\href {https://doi.org/10.1073/pnas.0307545100} {\bibfield  {journal} {\bibinfo  {journal} {Proceedings of the National Academy of Sciences}\ }\textbf {\bibinfo {volume} {101}},\ \bibinfo {pages} {5200} (\bibinfo {year} {2004})}\BibitemShut {NoStop}%
\bibitem [{\citenamefont {Barabási}\ \emph {et~al.}(2002)\citenamefont {Barabási}, \citenamefont {Jeong}, \citenamefont {Néda}, \citenamefont {Ravasz}, \citenamefont {Schubert},\ and\ \citenamefont {Vicsek}}]{barabasi_evolution_2002}%
  \BibitemOpen
  \bibfield  {author} {\bibinfo {author} {\bibfnamefont {A.~L.}\ \bibnamefont {Barabási}}, \bibinfo {author} {\bibfnamefont {H.}~\bibnamefont {Jeong}}, \bibinfo {author} {\bibfnamefont {Z.}~\bibnamefont {Néda}}, \bibinfo {author} {\bibfnamefont {E.}~\bibnamefont {Ravasz}}, \bibinfo {author} {\bibfnamefont {A.}~\bibnamefont {Schubert}},\ and\ \bibinfo {author} {\bibfnamefont {T.}~\bibnamefont {Vicsek}},\ }\href {https://doi.org/10.1016/S0378-4371(02)00736-7} {\bibfield  {journal} {\bibinfo  {journal} {Physica A: Statistical Mechanics and its Applications}\ }\textbf {\bibinfo {volume} {311}},\ \bibinfo {pages} {590} (\bibinfo {year} {2002})}\BibitemShut {NoStop}%
\bibitem [{\citenamefont {Lee}\ \emph {et~al.}(2010)\citenamefont {Lee}, \citenamefont {Goh}, \citenamefont {Kahng},\ and\ \citenamefont {Kim}}]{lee_complete_2010}%
  \BibitemOpen
  \bibfield  {author} {\bibinfo {author} {\bibfnamefont {D.}~\bibnamefont {Lee}}, \bibinfo {author} {\bibfnamefont {K.-I.}\ \bibnamefont {Goh}}, \bibinfo {author} {\bibfnamefont {B.}~\bibnamefont {Kahng}},\ and\ \bibinfo {author} {\bibfnamefont {D.}~\bibnamefont {Kim}},\ }\href {https://doi.org/10.1103/PhysRevE.82.026112} {\bibfield  {journal} {\bibinfo  {journal} {Physical Review E}\ }\textbf {\bibinfo {volume} {82}},\ \bibinfo {pages} {026112} (\bibinfo {year} {2010})}\BibitemShut {NoStop}%
\bibitem [{\citenamefont {Lee}\ \emph {et~al.}(2021{\natexlab{a}})\citenamefont {Lee}, \citenamefont {Lee}, \citenamefont {Oh}, \citenamefont {Lee},\ and\ \citenamefont {Kahng}}]{lee_homological_2021}%
  \BibitemOpen
  \bibfield  {author} {\bibinfo {author} {\bibfnamefont {Y.}~\bibnamefont {Lee}}, \bibinfo {author} {\bibfnamefont {J.}~\bibnamefont {Lee}}, \bibinfo {author} {\bibfnamefont {S.~M.}\ \bibnamefont {Oh}}, \bibinfo {author} {\bibfnamefont {D.}~\bibnamefont {Lee}},\ and\ \bibinfo {author} {\bibfnamefont {B.}~\bibnamefont {Kahng}},\ }\href {https://doi.org/10.1063/5.0047608} {\bibfield  {journal} {\bibinfo  {journal} {Chaos: An Interdisciplinary Journal of Nonlinear Science}\ }\textbf {\bibinfo {volume} {31}},\ \bibinfo {pages} {041102} (\bibinfo {year} {2021}{\natexlab{a}})}\BibitemShut {NoStop}%
\bibitem [{\citenamefont {Lee}\ \emph {et~al.}(2021{\natexlab{b}})\citenamefont {Lee}, \citenamefont {Lee}, \citenamefont {Oh},\ and\ \citenamefont {Kahng}}]{lee_betweenness_2021}%
  \BibitemOpen
  \bibfield  {author} {\bibinfo {author} {\bibfnamefont {J.}~\bibnamefont {Lee}}, \bibinfo {author} {\bibfnamefont {Y.}~\bibnamefont {Lee}}, \bibinfo {author} {\bibfnamefont {S.~M.}\ \bibnamefont {Oh}},\ and\ \bibinfo {author} {\bibfnamefont {B.}~\bibnamefont {Kahng}},\ }\href {https://doi.org/10.1063/5.0056683} {\bibfield  {journal} {\bibinfo  {journal} {Chaos: An Interdisciplinary Journal of Nonlinear Science}\ }\textbf {\bibinfo {volume} {31}},\ \bibinfo {pages} {061108} (\bibinfo {year} {2021}{\natexlab{b}})}\BibitemShut {NoStop}%
\bibitem [{\citenamefont {Oh}\ \emph {et~al.}(2021)\citenamefont {Oh}, \citenamefont {Lee}, \citenamefont {Lee},\ and\ \citenamefont {Kahng}}]{oh_emergence_2021}%
  \BibitemOpen
  \bibfield  {author} {\bibinfo {author} {\bibfnamefont {S.~M.}\ \bibnamefont {Oh}}, \bibinfo {author} {\bibfnamefont {Y.}~\bibnamefont {Lee}}, \bibinfo {author} {\bibfnamefont {J.}~\bibnamefont {Lee}},\ and\ \bibinfo {author} {\bibfnamefont {B.}~\bibnamefont {Kahng}},\ }\href {https://doi.org/10.1088/1742-5468/ac1667} {\bibfield  {journal} {\bibinfo  {journal} {Journal of Statistical Mechanics: Theory and Experiment}\ }\textbf {\bibinfo {volume} {2021}},\ \bibinfo {pages} {083218} (\bibinfo {year} {2021})}\BibitemShut {NoStop}%
\bibitem [{\citenamefont {Petri}\ \emph {et~al.}(2014)\citenamefont {Petri}, \citenamefont {Expert}, \citenamefont {Turkheimer}, \citenamefont {Carhart-Harris}, \citenamefont {Nutt}, \citenamefont {Hellyer},\ and\ \citenamefont {Vaccarino}}]{petri_homological_2014}%
  \BibitemOpen
  \bibfield  {author} {\bibinfo {author} {\bibfnamefont {G.}~\bibnamefont {Petri}}, \bibinfo {author} {\bibfnamefont {P.}~\bibnamefont {Expert}}, \bibinfo {author} {\bibfnamefont {F.}~\bibnamefont {Turkheimer}}, \bibinfo {author} {\bibfnamefont {R.}~\bibnamefont {Carhart-Harris}}, \bibinfo {author} {\bibfnamefont {D.}~\bibnamefont {Nutt}}, \bibinfo {author} {\bibfnamefont {P.~J.}\ \bibnamefont {Hellyer}},\ and\ \bibinfo {author} {\bibfnamefont {F.}~\bibnamefont {Vaccarino}},\ }\href {https://doi.org/10.1098/rsif.2014.0873} {\bibfield  {journal} {\bibinfo  {journal} {Journal of The Royal Society Interface}\ }\textbf {\bibinfo {volume} {11}},\ \bibinfo {pages} {20140873} (\bibinfo {year} {2014})}\BibitemShut {NoStop}%
\bibitem [{\citenamefont {Giusti}\ \emph {et~al.}(2016)\citenamefont {Giusti}, \citenamefont {Ghrist},\ and\ \citenamefont {Bassett}}]{giusti_twos_2016}%
  \BibitemOpen
  \bibfield  {author} {\bibinfo {author} {\bibfnamefont {C.}~\bibnamefont {Giusti}}, \bibinfo {author} {\bibfnamefont {R.}~\bibnamefont {Ghrist}},\ and\ \bibinfo {author} {\bibfnamefont {D.~S.}\ \bibnamefont {Bassett}},\ }\href {https://doi.org/10.1007/s10827-016-0608-6} {\bibfield  {journal} {\bibinfo  {journal} {Journal of Computational Neuroscience}\ }\textbf {\bibinfo {volume} {41}},\ \bibinfo {pages} {1} (\bibinfo {year} {2016})}\BibitemShut {NoStop}%
\bibitem [{\citenamefont {Sizemore}\ \emph {et~al.}(2018)\citenamefont {Sizemore}, \citenamefont {Giusti}, \citenamefont {Kahn}, \citenamefont {Vettel}, \citenamefont {Betzel},\ and\ \citenamefont {Bassett}}]{sizemore_cliques_2018}%
  \BibitemOpen
  \bibfield  {author} {\bibinfo {author} {\bibfnamefont {A.~E.}\ \bibnamefont {Sizemore}}, \bibinfo {author} {\bibfnamefont {C.}~\bibnamefont {Giusti}}, \bibinfo {author} {\bibfnamefont {A.}~\bibnamefont {Kahn}}, \bibinfo {author} {\bibfnamefont {J.~M.}\ \bibnamefont {Vettel}}, \bibinfo {author} {\bibfnamefont {R.~F.}\ \bibnamefont {Betzel}},\ and\ \bibinfo {author} {\bibfnamefont {D.~S.}\ \bibnamefont {Bassett}},\ }\href {https://doi.org/10.1007/s10827-017-0672-6} {\bibfield  {journal} {\bibinfo  {journal} {Journal of Computational Neuroscience}\ }\textbf {\bibinfo {volume} {44}},\ \bibinfo {pages} {115} (\bibinfo {year} {2018})}\BibitemShut {NoStop}%
\bibitem [{\citenamefont {Bianconi}(2015)}]{bianconi_interdisciplinary_2015}%
  \BibitemOpen
  \bibfield  {author} {\bibinfo {author} {\bibfnamefont {G.}~\bibnamefont {Bianconi}},\ }\href {https://doi.org/10.1209/0295-5075/111/56001} {\bibfield  {journal} {\bibinfo  {journal} {{EPL} (Europhysics Letters)}\ }\textbf {\bibinfo {volume} {111}},\ \bibinfo {pages} {56001} (\bibinfo {year} {2015})}\BibitemShut {NoStop}%
\bibitem [{\citenamefont {Iacopini}\ \emph {et~al.}(2019)\citenamefont {Iacopini}, \citenamefont {Petri}, \citenamefont {Barrat},\ and\ \citenamefont {Latora}}]{iacopini_simplicial_2019}%
  \BibitemOpen
  \bibfield  {author} {\bibinfo {author} {\bibfnamefont {I.}~\bibnamefont {Iacopini}}, \bibinfo {author} {\bibfnamefont {G.}~\bibnamefont {Petri}}, \bibinfo {author} {\bibfnamefont {A.}~\bibnamefont {Barrat}},\ and\ \bibinfo {author} {\bibfnamefont {V.}~\bibnamefont {Latora}},\ }\href {https://doi.org/10.1038/s41467-019-10431-6} {\bibfield  {journal} {\bibinfo  {journal} {Nature Communications}\ }\textbf {\bibinfo {volume} {10}},\ \bibinfo {pages} {2485} (\bibinfo {year} {2019})}\BibitemShut {NoStop}%
\bibitem [{\citenamefont {Jhun}\ \emph {et~al.}(2019)\citenamefont {Jhun}, \citenamefont {Jo},\ and\ \citenamefont {Kahng}}]{jhun_simplicial_2019}%
  \BibitemOpen
  \bibfield  {author} {\bibinfo {author} {\bibfnamefont {B.}~\bibnamefont {Jhun}}, \bibinfo {author} {\bibfnamefont {M.}~\bibnamefont {Jo}},\ and\ \bibinfo {author} {\bibfnamefont {B.}~\bibnamefont {Kahng}},\ }\href {https://doi.org/10.1088/1742-5468/ab5367} {\bibfield  {journal} {\bibinfo  {journal} {Journal of Statistical Mechanics: Theory and Experiment}\ }\textbf {\bibinfo {volume} {2019}},\ \bibinfo {pages} {123207} (\bibinfo {year} {2019})}\BibitemShut {NoStop}%
\bibitem [{\citenamefont {de~Arruda}\ \emph {et~al.}(2020)\citenamefont {de~Arruda}, \citenamefont {Petri},\ and\ \citenamefont {Moreno}}]{de_arruda_social_2020}%
  \BibitemOpen
  \bibfield  {author} {\bibinfo {author} {\bibfnamefont {G.~F.}\ \bibnamefont {de~Arruda}}, \bibinfo {author} {\bibfnamefont {G.}~\bibnamefont {Petri}},\ and\ \bibinfo {author} {\bibfnamefont {Y.}~\bibnamefont {Moreno}},\ }\href {https://doi.org/10.1103/PhysRevResearch.2.023032} {\bibfield  {journal} {\bibinfo  {journal} {Physical Review Research}\ }\textbf {\bibinfo {volume} {2}},\ \bibinfo {pages} {023032} (\bibinfo {year} {2020})}\BibitemShut {NoStop}%
\bibitem [{\citenamefont {Dorogovtsev}\ \emph {et~al.}(2000)\citenamefont {Dorogovtsev}, \citenamefont {Mendes},\ and\ \citenamefont {Samukhin}}]{dorogovtsev_structure_2000}%
  \BibitemOpen
  \bibfield  {author} {\bibinfo {author} {\bibfnamefont {S.~N.}\ \bibnamefont {Dorogovtsev}}, \bibinfo {author} {\bibfnamefont {J.~F.~F.}\ \bibnamefont {Mendes}},\ and\ \bibinfo {author} {\bibfnamefont {A.~N.}\ \bibnamefont {Samukhin}},\ }\href {https://doi.org/10.1103/PhysRevLett.85.4633} {\bibfield  {journal} {\bibinfo  {journal} {Physical Review Letters}\ }\textbf {\bibinfo {volume} {85}},\ \bibinfo {pages} {4633} (\bibinfo {year} {2000})}\BibitemShut {NoStop}%
\bibitem [{\citenamefont {Callaway}\ \emph {et~al.}(2001)\citenamefont {Callaway}, \citenamefont {Hopcroft}, \citenamefont {Kleinberg}, \citenamefont {Newman},\ and\ \citenamefont {Strogatz}}]{callaway_are_2001}%
  \BibitemOpen
  \bibfield  {author} {\bibinfo {author} {\bibfnamefont {D.~S.}\ \bibnamefont {Callaway}}, \bibinfo {author} {\bibfnamefont {J.~E.}\ \bibnamefont {Hopcroft}}, \bibinfo {author} {\bibfnamefont {J.~M.}\ \bibnamefont {Kleinberg}}, \bibinfo {author} {\bibfnamefont {M.~E.~J.}\ \bibnamefont {Newman}},\ and\ \bibinfo {author} {\bibfnamefont {S.~H.}\ \bibnamefont {Strogatz}},\ }\href {https://doi.org/10.1103/PhysRevE.64.041902} {\bibfield  {journal} {\bibinfo  {journal} {Physical Review E}\ }\textbf {\bibinfo {volume} {64}},\ \bibinfo {pages} {041902} (\bibinfo {year} {2001})}\BibitemShut {NoStop}%
\bibitem [{\citenamefont {Dorogovtsev}\ \emph {et~al.}(2001)\citenamefont {Dorogovtsev}, \citenamefont {Mendes},\ and\ \citenamefont {Samukhin}}]{dorogovtsev_anomalous_2001}%
  \BibitemOpen
  \bibfield  {author} {\bibinfo {author} {\bibfnamefont {S.~N.}\ \bibnamefont {Dorogovtsev}}, \bibinfo {author} {\bibfnamefont {J.~F.~F.}\ \bibnamefont {Mendes}},\ and\ \bibinfo {author} {\bibfnamefont {A.~N.}\ \bibnamefont {Samukhin}},\ }\href {https://doi.org/10.1103/PhysRevE.64.066110} {\bibfield  {journal} {\bibinfo  {journal} {Physical Review E}\ }\textbf {\bibinfo {volume} {64}},\ \bibinfo {pages} {066110} (\bibinfo {year} {2001})}\BibitemShut {NoStop}%
\bibitem [{\citenamefont {Dorogovtsev}\ and\ \citenamefont {Mendes}(2002)}]{dorogovtsev_evolution_2002}%
  \BibitemOpen
  \bibfield  {author} {\bibinfo {author} {\bibfnamefont {S.~N.}\ \bibnamefont {Dorogovtsev}}\ and\ \bibinfo {author} {\bibfnamefont {J.~F.~F.}\ \bibnamefont {Mendes}},\ }\href {https://doi.org/10.1080/00018730110112519} {\bibfield  {journal} {\bibinfo  {journal} {Advances in Physics}\ }\textbf {\bibinfo {volume} {51}},\ \bibinfo {pages} {1079} (\bibinfo {year} {2002})}\BibitemShut {NoStop}%
\end{thebibliography}%

\end{document}